\documentclass[english,aps, manuscript]{revtex4}
\usepackage[T1]{fontenc}
\usepackage[latin1]{inputenc}
\usepackage{amsmath}

\makeatletter

\makeatletter
\usepackage{color}

\makeatother

\usepackage{babel}
\makeatother
\begin{document}

\title{Transitions Between Flux Vacua}

\author{S. P. de Alwis\protect{}}

\affiliation{Perimeter Institute, 31 Caroline Street N., Waterloo, ON N2L 2Y5,
Canada}

\affiliation{Department of Physics, University of Colorado, Box 390, Boulder,
CO 80309.\\
 \\
 \texttt{e-mail: dealwis@pizero.colorado.edu} PACS : 11.25. -w, 98.80.-k}

\begin{abstract}
A dynamical description of the transitions between different backgrounds
requires the existence of a background independent action which propagates
the correct number of degrees of freedom and couples bulk supergravity
to certain higher dimensional branes. We present classical equations
for configurations that separate the world into regions with different
flux parameters etc. and discuss the difficulties of trying to construct
an action that describes the transitions between them within the framework
of supergravity. 
\end{abstract}
\maketitle
\vfill{}

\eject

\section{Introduction}

The realization that the solutions to the equations of string theory
can give in four dimensions a large multiplicity of vacua (called
the landscape) has led to much discussion as to whether there is a
dynamical selection principle that picks one or a class of these vacua,
or whether we simply find ourselves living in a universe where observers
such as ourselves can exist. The former consists principally of arguments
from quantum cosmology and seems to apply only to closed universes.
In any case it is not clear that it can help resolve the cosmological
constant problem. The latter goes under the name of the Anthropic
Principle and it is not clear whether it is little more than a tautology.
At best it may help resolve the so-called cosmic coincidence problem.
In any case both approaches become meaningful only within a theory
in which there is a mechanism by which the different universes can
be realized.

In this paper we will discuss dynamical processes in the landscape,
by which transitions between vacua with different flux quantum numbers
and different numbers of branes can take place \cite{Brown:1988kg},\cite{Bousso:2000xa},\cite{Feng:2000if}.
Of course the landscape will also consist of different compactification
manifolds. Even if we restrict ourselves to Calabi-Yau compactifications
there are of the order of $10^{5}$ manifolds with different numbers
of two and three cycles. We do know that through conifold transitions
one can change these topological numbers, but it is not clear that
there is a dynamical process which describes this in string theory
\footnote{For a discussion of this in a mini-superspace (cosmological) context
see \cite{Mohaupt:2004pr}\cite{Lukas:2004du}\cite{Mohaupt:2005pa}. %
}. On the other hand it is widely believed that such a process does
exist for changing flux quanta and the number of D-branes. Here we
will be concerned with this.

We will discuss only type IIB and IIA compactifications. Moduli stabilization
issues in the heterotic and M-theory cases are less well understood.
We will show that there are indeed classical configurations that describe
a variety of transitions. However the processes involve the nucleation
of various higher dimensional branes. These are branes which are magnetically
coupled to the fields in the bulk theory. In the case of type IIB
these will be 5-branes (either NSNS or RR) and in the case of IIA
they will be 6- and 8-branes. The process however is essentially quantum
mechanical. The existence of a classical configuration which divides
space into two regions separated by a domain wall (i.e. the nucleated
higher dimensional brane) by itself does not mean that the process
of nucleation can take place. For this, at the very least, one should
be able to construct an action for the bulk fields and the brane.
We attempt to do this here in the low energy (i.e. supergravity) limit
of string theory.

In type IIB we have an immediate problem in that the self-duality
condition for the five-form field does not permit us to write a (Lorentz
invariant) action. This problem is usually addressed by imposing the
self-duality condition at the level of the equations of motion. We
adopt this procedure tentatively (even though it is not a proper action
for a quantum theory) and write down the coupling to three and five-D-branes
and orientifold planes. We discuss in detail the classical equations
and the transitions between flux vacua. We show how to construct a
bulk action magnetically coupled to the five brane by using the phenomenon
of anomaly inflow. However at the end of the day it turns out that
this action is inconsistent with the self duality constraint. In other
words although in the absence of the five-brane there is no inconsistency,
once the five brane is introduced the equations of motion and Bianchi
identities are inconsistent with the self-duality constraint.

It might be thought that an alternative procedure would be to consider
a non-Lorentz invariant action for type IIB since one is breaking
the (10 dimensional) Lorentz invariance anyway. However the self-duality
condition involves the metric, and an implementation of it which picks
out the four directions of the external space will be dependent on
the background geometry of the space. This is more than just a topological
restriction to work in spaces of the form $M_{4}\times X_{6}$, it
depends also on the metric on $M_{4}$. This is manifestly unsuited
to a discussion of processes which change the background - in particular
the cosmological constant is changed by these processes. It can hardly
be over-emphasized that such a discussion must necessarily be made
only with a background independent formulation %
\footnote{A discussion of IIB coupling to branes using the PST formalism, which
can be used if the manifold is topologically trivial, is given in
\cite{Cariglia:2004ez}. A discussion of a non-Lorentz invariant formalism
that is valid in topologically non-trivial manifolds is given in \cite{Belov:2006jd}.%
}.

Since in the IIB case this problem might be attributed to the lack
of a proper action even in the absence of the nucleated five-brane,
it is reasonable to expect that there is no problem in IIA, where
there is no such self-duality constraint. We discuss in detail the
classical equations (Bianchi identities) that lead to flux changing
processes resulting from the nucleation of D6 and D8 branes. After
some manipulations we find that it is indeed possible to construct
a bulk action coupled to 6 and 8- branes for the massive IIA theory
\footnote{This corrects a statement in an earlier draft of this paper. I thank
Dima Belov and Greg Moore for suggestions that led me to find the
relevant extra terms in the bulk action.%
} that is gauge invariant. However this action is explicitly dependent
on the background fluxes.

Now it could still be the case that although there is no semi-classical
description, the complete formulation of string theory contains a
quantum mechanical description of these processes. On the other hand
it seems to us that since it is assumed that classical flux compactification
arguments that lead to moduli stabilization and the landscape, survive
in the full string theory, the absence of a background independent
semi-classical description of transitions might also survive. If that
is so it would mean that each point in the landscape is simply a model
and there is no need to ascribe any degree of reality to any of them
except the one (if it exists) that contains the standard model with
a tiny cosmological constant. On the other hand given that a classical
configuration that divides space into two different regions does exist,
it is arguable that the technical point highlighted in this paper
will be overcome in the full string theory, and that a quantum process
of brane nucleation of these higher dimensional branes is in fact
allowed.

\section{String theory processes}

\subsection{Type IIB}

A concrete framework in which such a discussion can be made is that
of Kachru et al. \cite{Kachru:2002gs} (see also\cite{Frey:2003dm}).
Here the nucleation of a NS5 branes that take the form of a $S_{2\,\,}$
bubble wall in non compact four dimensions and sweeps out an $S_{3}$
in the six dimensional Calabi-Yau manifold $X$, is discussed within
the GKP context. The flux constraint coming from the Bianchi identity
for the five-form flux is \begin{equation}
\frac{\chi}{24}=N_{3}-\bar{N}_{3}-\frac{1}{2\kappa_{10}^{2}T_{3}}\int_{X}H_{3}\wedge F_{3}\label{chiX}\end{equation}
 where $H_{3}(F_{3})$ are NSNS(RR) three form fluxes $\chi$ is the
Euler character of the associated Calabi-Yau 4-fold in the F-theory
context (alternatively the LHS is the contribution of orientifold
planes in the purely six dimensional context). Assuming that that
there is only one pair of crossed fluxes (setting $2\pi\alpha'=1$)\begin{equation}
\int_{A}F_{3}=2\pi M,\,\int_{B}H_{3}=-2\pi K\label{fluxes}\end{equation}
 (\ref{chiX}) becomes \begin{equation}
\frac{\chi}{24}=N_{3}-\bar{N}_{3}-KM\label{chiX2}\end{equation}

The argument of \cite{Kachru:2002gs} is that the nucleation of NS
five-branes will change the flux on the cycle dual to the one that's
wrapped by the five brane $K\rightarrow K\pm1$ so that by the above
equation the net number of D3 branes will change by M. So by such
a transition one would expect a string theory realization of the BT
process. Obviously this process explores only points on the landscape
which all belong to a given F-theory compactification (or CY orientifold)
but it is still important to establish whether this can actually take
place and be described in semi-classical terms.

Everything that was said above will have an S-dual counterpart. Clearly
a dual process would be one in which a D5 brane (wrapping the B cycle)
is nucleated so that $M\rightarrow M\pm1$ with the number of D3 branes
changing by $\mp K$. Since D5 brane actions are better understood
than NS five branes we will focus on this rather than its S-dual.
Our aim is to see whether an action which describes these D3 brane
configurations interpolated by a D5 brane exists.

Since the dilaton-axion system is irrelevant for our considerations
let us freeze them by putting $e^{\phi}=1$ and $c_{0}=0.$ The form
of the bulk action (for the gauge fields) is \begin{eqnarray}
S_{IIB} & = & \frac{1}{(2\pi)^{3}}\int_{M_{10}}[-\frac{1}{4}F_{5}\wedge*F_{5}-\frac{1}{2}H_{3}\wedge*H_{3}-\frac{1}{2}F_{3}\wedge*F_{3}]\nonumber \\
 &  & +\frac{1}{2}\frac{1}{(2\pi)^{3}}\int_{D_{11}}F_{5}\wedge F_{3}\wedge H_{3}\label{Jbulkaction}\end{eqnarray}
 The last (topological) term is integrated over a eleven disc whose
boundary is the ten manifold $M_{10}.$ In the absence of sources
the integrand is closed due to the Bianchi identities (given below)
so that the integral is independent of the particular disc over which
the definition of the fields is extended, provided also that the integral
over an arbitrary closed 11-manifold is ($(2\pi)^{4}$ times) an even
integer (see for example \cite{Witten:1996hc}). The field strengths
satisfy the Bianchi identities\begin{eqnarray}
dH_{3} & = & 0,\, dF_{3}=0\label{Bi1}\\
dF_{5} & = & H_{3}\wedge F_{3},\label{Bi2}\end{eqnarray}
 which are solved locally by\begin{eqnarray}
H_{3}=dB_{2} &  & F_{3}=dC_{2}\label{H3F3}\\
F_{5} & = & dC_{4}-H_{3}\wedge C_{2}.\label{F5}\end{eqnarray}

In the absence of sources or non-trivial $F_{5}$ flux the last integral
in the action can be written as\[
\int_{M_{10}}C_{4}\wedge F_{3}\wedge H_{3}.\]
 The equations of motion and the self-duality constraint are\begin{eqnarray}
d*F_{3}=F_{5}\wedge H_{3}, &  & d*H_{3}=F_{3}\wedge F_{5},\label{eom}\\
F_{5} & = & *F_{5}.\label{sd}\end{eqnarray}
 Note that the local solution to the Bianchi identities are RR fields
which are related to the ones given in \cite{Polchinski:1998rr} by
the substitution $C_{4}\rightarrow C_{4}+\frac{1}{2}B_{2}\wedge C_{2}$
\footnote{Note that the extra topological term which arises under this substitution
does not contributed to the equation of motion and may be omitted.%
}. $ $ The gauge transformations are as follows. \begin{eqnarray}
\delta B_{2} & = & d\Lambda_{1},\,\delta C_{4}=0,\label{JBtrans}\\
\delta C_{2} & = & d\tilde{\Lambda}_{1},\,\delta C_{4}=-H_{3}\wedge\tilde{\Lambda}_{1},\label{JCtrans}\\
\delta C_{4} & = & d\tilde{\Lambda}_{3}.\label{JC4trans}\end{eqnarray}
 The WZNW part of the D3 brane action is then the usual one\begin{equation}
I_{3}=\frac{\mu_{3}}{(2\pi)^{3}}\int_{W_{4}}[C_{4}-C_{2}\wedge{\mathcal{F}}_{2}],\label{I3}\end{equation}
 where \begin{equation}
{\mathcal{F}}_{2}=B_{2}+f_{2}.\label{calF}\end{equation}
 Here $f_{2}$ is the world volume gauge field strength which under
the gauge transformation (\ref{JBtrans}) transforms as $\delta f_{2}=d\Lambda$,
and $\mu_{3}=(2\pi)^{2}q_{3}\,(q_{3}=\pm1,0$ for D3, an anti-D3 or
no D3). (\ref{I3}) is invariant under (\ref{JBtrans},\ref{JCtrans},\ref{JC4trans}).
If this is coupled to the bulk action (i.e. $q_{3}=\pm1$) then the
Bianchi identity (\ref{Bi2}) is modified. The $C_{4}$ equation of
motion for the total action $S_{IIB}+I_{3}$ and self-duality of $F_{5}$
gives\begin{equation}
dF_{5}=H_{3}\wedge F_{3}-2\mu_{3}\delta_{6}(M_{10}\rightarrow W_{4})+\ldots.\label{modBi2}\end{equation}
 where the ellipses denote the contributions of orientifold planes
(or D7 brane contributions in the case of F-theory). However it was
argued in footnote 6 of GKP \cite{Giddings:2001yu} that in deriving
this from the action for the D3 brane, the topological term needs
to be taken as half the value given in (\ref{I3}). This is related
to the fact that because of the self duality of the D3 brane the electric
and magnetic couplings are identical. Thus in GKP the above relation
is written as\begin{equation}
dF_{5}=H_{3}\wedge F_{3}-\mu_{3}\delta_{6}(M_{10}\rightarrow W_{4})+\ldots.\label{GKPBi}\end{equation}
 This formula in fact plays a crucial role in the GKP analysis. It
should be mentioned here that (as emphasized by GKP) the action is
still supposed to be taken to be the sum of the type IIB bulk action
and the D3 brane action (i.e. without any relative coefficient such
as a factor half). In particular GKP use the gravitational variation
of the DBI part of the action in their analysis and of course the
relation between the coefficient of that part (the tension) and the
charge $\mu_{3}$ is fixed by supersymmetry. In fact it is easy to
see that this is the equation that is consistent with the equation
of motion for the RR $C_{2}$ field. In the presence of the coupling
to the three brane (\ref{I3}) the first equation of (\ref{eom})
is modified to

\begin{equation}
d*F_{3}=F_{5}\wedge H_{3}+\mu_{3}{\mathcal{F}}_{2}\wedge\delta_{6}(M_{10}\rightarrow W_{4})\label{eommod}\end{equation}

Requiring consistency with $d^{2}=0$ gives\[
0=dF_{5}\wedge H_{3}+\mu_{3}H_{3}\wedge\delta_{6}(M_{10}\rightarrow W_{4})\]
 where we've used the fact that $d{\cal F}{}_{2}=H_{3}$. Clearly
this is consistent only with (\ref{GKPBi}). This justifies the choice
made by GKP. What goes wrong with (\ref{modBi2}) is that it is really
an equation of motion i.e. gives us from the Lagrangian $d*F_{5}$,
which does not conflict with (\ref{eommod}) by itself. A (gauge invariant)
Lagrangian cannot give inconsistent equations of motion! It is really
the imposition of the self-duality constraint (\ref{sd}) that creates
a problem. Unfortunately this means that there is no (Lorentz invariant)
Lagrangian formulation of the bulk action coupled to D3 branes that
is consistent with self duality of the five form, even if the latter
is imposed by hand at the level of the equations of motion.

Now let us try to couple D5 branes to the bulk IIB action. First of
all it should be emphasized that there is no way of coupling the six-form
field $C_{6}$ to the bulk action. There are two related problems
in doing this. One could proceed as usual to dualize by introducing
a Lagrange multiplier to switch the Bianchi identity and the EOM of
the three form field $C_{2}$ effectively replacing $\pm*F_{3}\rightarrow F_{7}=dC_{6}+\ldots$.
If one just had the kinetic term for $C_{6}$ then indeed this would
make sense and writing the coupled action as\[
\frac{1}{(2\pi)^{3}}\int_{M_{10}}[-\frac{1}{2}F_{7}\wedge*F_{7}]+\frac{\mu_{5}}{(2\pi)^{3}}\int_{W_{6}}C_{6},\]
 we would get the the equation of motion for $C_{6}$ which would
be the Bianchi identity for $F_{3}$- namely,\begin{equation}
dF_{3}=\pm d*F_{7}=-\mu_{5}\delta_{4}(M_{10}\rightarrow W_{6}).\label{C6Bi}\end{equation}
 However the problem is that this dualization cannot be carried out
in the full IIB action. The $C_{2}$ form cannot be removed from the
action since it occurs explicitly (i.e. not just through its curvature)
in the (local) solution to the Bianchi identity for $\tilde{F}_{5}$.
Furthermore both $C_{4}$ and $C_{2}$ occur in the D5 brane action
for precisely the same reason - namely gauge invariance. In fact a
similar situation is encountered in trying to dualize the M-theory
action to couple M5 branes and for a detailed discussion of the problems
that one encounters in that case see \cite{Witten:1996hc}\cite{deAlwis:1997gq}.
The resolution is the same. We drop the higher form terms entirely
from the action and simply impose the Bianchi identity. The dropped
terms will then reappear in the bulk action as Dirac string terms.
Let us see how this works in detail.

The topological terms in the action for a D5 brane, in the form in
which can couple to the above bulk action are,\begin{equation}
I_{5}=\frac{\mu_{5}}{(2\pi)^{3}}\int_{W_{6}}[-C_{4}\wedge{\mathcal{F}}_{2}+\frac{1}{2!}C_{2}\wedge{\mathcal{F}}_{2}^{2}]\label{I5}\end{equation}
 Here $\mu_{5}=2\pi q_{5}$ ($q_{5}=\pm1,0$ for D5, a D-bar 5 or
no D5) and ${\mathcal{F}}_{2}=B_{2}+f_{2},$ as before except that
now the field $f_{2}$ lives on the world volume $W_{6}$. Note that
we have omitted the $\int C_{6}$ term since this gauge field is absent
from the bulk action. Instead %
\footnote{This method was for coupling higher dimensional branes to the bulk
was discussed in the context of M-theory in \cite{deAlwis:1997gq,deAlwis:1997rd}.%
} we require that the Bianchi identity for $C_{3}$ is changed from
(\ref{Bi1}) to \begin{equation}
dF_{3}=2\lambda^{-1}\mu_{5}\delta_{4}(M_{10}\rightarrow W_{6})\label{Bimod}\end{equation}

The coefficient in front of the delta function in the above will be
fixed by gauge invariance. Both the topological term in the bulk action
as well as that in the five brane action are now separately anomalous
under the gauge transformation (\ref{JC4trans}) which leaves the
bulk action (\ref{Jbulkaction}) invariant in the absence of sources,
i.e. with zero on the RHS of (\ref{Bimod}). In the eleven dimensional
form of the topological term, the presence of the five brane means
that the integrand of the topological term (the second line of (\ref{Jbulkaction})
is ambiguous. We choose to fix this ambiguity by imposing gauge invariance
of the combined system bulk plus brane. \begin{equation}
S_{top}=\frac{1}{(2\pi)^{3}}\int_{M_{10}}[\frac{\lambda}{2}C_{4}\wedge F_{3}\wedge H_{3}+\frac{1-\lambda}{2}C_{4}\wedge F'_{3}\wedge H_{3}]\label{stopIIB}\end{equation}
 where $F_{3}$ obeys the Bianchi identity (\ref{Bimod}) and $dF'_{3}=0$
so that locally $F_{3}'=dC_{2}$. In fact we may write $F_{3}=F'_{3}+\theta_{3}$
where $\theta_{3}$ (the Dirac string term) may be defined as the
coexact solution to (\ref{Bimod}). Given this the second term is
gauge invariant but because of (\ref{Bimod}) the first gives\begin{equation}
\delta S=\frac{1}{(2\pi)^{3}}\frac{\lambda}{2}\int\tilde{\Lambda}_{3}\wedge dF_{3}\wedge H_{3}=\frac{\mu_{5}}{(2\pi)^{3}}\int_{W_{6}}\tilde{\Lambda}_{3}\wedge H_{3}.\label{deltaS}\end{equation}

Let us now look at the gauge variation of the five brane action (\ref{I5}).
First note that under the $B_{2}$ gauge variation $\delta f_{2}=-d\Lambda_{1}$
so that ${\mathcal{F}}_{2}$ is gauge invariant. So we have\begin{eqnarray}
\delta I_{5} & = & \frac{\mu_{5}}{(2\pi)^{3}}\int_{W_{6}}[-(d\tilde{\Lambda}_{3}-H_{3}\wedge\tilde{\Lambda}_{1})\wedge{\mathcal{F}}_{2}+\frac{1}{2}d\tilde{\Lambda}_{1}\wedge{\mathcal{F}}_{2}^{2}]\nonumber \\
 & = & -\frac{\mu_{5}}{(2\pi)^{3}}\int_{W_{6}}\tilde{\Lambda}_{3}\wedge H_{3}\label{deltaI5}\end{eqnarray}
 In the last line above we used the formula\begin{equation}
d{\cal F}{}_{2}=dB_{2}+dF_{2}=H_{3}\label{Bi3}\end{equation}
 where in the last equality we used the Bianchi identity for the gauge
field strength on the five-brane world volume. Thus the combined action
$S+I_{5}$ of bulk IIB SUGRA plus the six brane is gauge invariant
by the phenomenon of anomaly inflow.

The coefficient $\lambda$ in (\ref{Bimod}) can now be fixed by requiring
the consistency of the Bianchi identity for the five form field when
the bulk is coupled to the five brane action (\ref{I5}). Again as
was the case in the coupling of the D3 brane, the $C_{4}$ equation
of motion (and the self-duality constraint) gives the wrong answer
by a factor of two. The correct result is\begin{equation}
dF_{5}=H_{3}\wedge F_{3}+\mu_{5}{\cal F}{}_{2}\delta_{4}(M_{10}\rightarrow W_{6})+\ldots\label{mod2Bi2}\end{equation}
 since it must agree with (\ref{GKPBi}) when the D5 brane contains
a dissolved D3 brane. In detail this may be seen by comparing the
two equations after integrating over the six manifold and giving a
unit magnetic flux$\int_{S_{2}}f_{2}=-2\pi$ and using the relation
$\mu_{3}=2\pi\mu_{5}$. Then consistency with $d^{2}=0$ (and the
use of the formula $d{\cal F}{}_{2}=H_{3}$) gives $\lambda=2$ in
(\ref{Bimod}). So the Bianchi identity is

\begin{equation}
dF_{3}=\mu_{5}\delta_{4}(M_{10}\rightarrow W_{6})\label{F3BI}\end{equation}
 which is the same as (\ref{C6Bi}) if we choose the negative sign
in the first equality.

Suppose that the nucleated D5 brane wraps the spatial directions $S_{2}\times S_{3}^{B}$
where the first factor is in the non-compact space and the last is
a three (B) cycle in X. Integrating (\ref{F3BI}) over $R\times S_{3}^{A}$
where the factor $R$ is a radial direction going from inside the
bubble (whose wall is an $S_{2}$) to the outside and $S_{3}^{A}$
is the cycle dual to $S_{3}^{B}$, we get\begin{equation}
\int_{R\times S_{3}^{A}}dF_{3}=\Delta\int_{S_{3}^{A}}F_{3}=\pm(2\pi)\label{DF3}\end{equation}
 i.e. the RR flux (\ref{fluxes}) changes by one unit $M\rightarrow M\pm1$
as one goes from inside the bubble to outside or vice-versa.

Now let us consider the topological terms for the D3 branes. These
branes have boundaries which are attached to the D5 brane. Now since
the world volume has boundaries we have from (\ref{I3},\ref{JCtrans},\ref{JC4trans})\begin{equation}
\delta I_{3}=\frac{\mu_{3}}{(2\pi)^{3}}\int_{\partial W_{4}}(\tilde{\Lambda}_{3}-\tilde{\Lambda}_{1}\wedge{\cal F}{}_{2}).\label{deltaI3}\end{equation}
 If the boundary of this D3 brane lies on the D5 brane there is effectively
a monopole on the D5 brane world volume wrapping an $S_{3}$ cycle
in $X$ so that $df_{2}\ne0$ on the D5 brane world volume. Thus there
is an uncanceled piece in the gauge transformation of $I_{5}$,\begin{equation}
\delta^{+}I_{5}=-\frac{\mu_{5}}{(2\pi)^{3}}\int_{W_{6}}(\tilde{\Lambda}_{3}-\tilde{\Lambda}_{1}\wedge{\cal F}{}_{2})\wedge df_{2}.\label{delta+I5}\end{equation}
 Now on the D5 brane we have\[
d{\mathcal{F}}_{2}=H_{3}+df_{2}\]
 Integrating this over the $S_{3}$ in X wrapped by the D5 brane and
using the fact that ${\mathcal{F}}_{2}$ (unlike $f_{2}$) is globally
defined we have\begin{equation}
\int_{S_{3}}df_{2}=-\int_{S_{3}}H_{3}=2\pi K,\, K\varepsilon{\mathcal{Z}}\label{Hflux}\end{equation}
 Thus the monopole equation is \begin{equation}
df_{2}=2\pi\sum_{i}q_{i}\delta_{3}(S_{3}\rightarrow i)\label{monopole}\end{equation}
 with\begin{equation}
\sum_{i}q_{i}\equiv N_{3}-\bar{N}_{3}=K\label{NNbar}\end{equation}
 where $N_{3}(\bar{N}_{3})$ are the number of D3 (D-bar3) branes
ending on this D5 brane. Substituting (\ref{monopole}) into (\ref{delta+I5})
and using $\mu_{5}=2\pi,\,\mu_{3}=(2\pi)^{2}$ we see that the gauge
variation of the D3 brane action is cancelled by anomaly inflow from
the D5 brane. Thus we have shown that the total action\begin{equation}
S_{T}=S_{IIB}+I_{5}+\sum_{i}I_{3i}\label{Stotal}\end{equation}
 is gauge invariant.

The modified Bianchi identity (replacing (\ref{GKPBi}) and (\ref{mod2Bi2})
is\begin{equation}
dF_{5}=H_{3}\wedge F_{3}+\mu_{5}{\mathcal{F}}_{2}\wedge\delta_{4}(M_{10}\rightarrow W_{6})-\sum_{i}\mu_{3}^{i}\delta_{6}(M_{10}\rightarrow W_{4}^{i})+\ldots\label{FinalBi}\end{equation}
 where again the ellipses denote the contribution of orientifold planes
(one may also have D3 branes that do not end on the D5 brane). The
relative coefficients of this Bianchi identity in fact cannot be altered.
As seen above the relative coefficients of the second and the third
term depends on the T-duality relation $\mu_{3}/\mu_{5}=2\pi$, and
the relative coefficient and sign between the first and second terms
is determined by gauge invariance. This may be checked %
\footnote{In checking this we need to use the formula $d\delta_{6}(M_{10}\rightarrow W_{4})=\delta_{7}(M\rightarrow\partial W_{4})$.
When the world volume has no boundary the right hand side is zero.%
} by using the formula $d^{2}=0$ and (\ref{F3BI}) where the latter
was fixed by the anomaly inflow argument. The point is that once the
action is defined to be (\ref{Stotal}) gauge invariance (and T-duality)
fixes the Bianchi identity, and it cannot be modified on the grounds
that the equation of motion for $C_{4}$ needs to be modified because
of self-duality. Integrating over the internal manifold X (and assuming
for simplicity that there is only the one pair of crossed fluxes and
only the D (Dbar) 3-branes that end on the D5 brane) we find (\ref{chiX2}).
As observed above as one goes from the outside to the inside of the
bubble $M\rightarrow M+1$, so that (using (\ref{NNbar}) we have
the (dual of the) phenomenon discussed in \cite{Kachru:2002gs} of
branes being replaced by flux and vice-versa.

Thus we might have had a representation in terms of an action for
the BT process with the self-duality imposed at the level of the equations
of motion, if not for the fact that the equation of motion for $C_{4}$
in the coupled brane-bulk action and the self duality condition lead
to a result that is inconsistent with the correct Bianchi identity.
As far as we are aware there is no Lorentz invariant formalism that
can fix this problem.

\subsection{Type IIA}

Flux compactification in type IIA string theory has been discussed
by \cite{Grimm:2004ua}\cite{Villadoro:2005cu}\cite{DeWolfe:2005uu}.
As in the previous subsection we will try to write down an action
that describes processes which describe transitions between different
flux vacua. Unlike the case of IIB there is no reason to expect a
problem in this since there is no self duality constraint.

The bulk action for the p-form fields is (again with $2\pi\alpha'=1$
and a frozen axio-dilaton)\begin{eqnarray}
S & = & -\frac{1}{2}\frac{1}{(2\pi)^{3}}[\int_{M_{10}}(F_{0}\wedge*F_{0}+F_{2}\wedge*F_{2}+H_{3}\wedge*H_{3}+F_{4}\wedge*F_{4})\nonumber \\
 &  & +\int_{D_{11}}F_{4}\wedge F_{4}\wedge H_{3}]\label{SIIA}\end{eqnarray}

As befits a topological term the last one in the above expression
is integrated over a disc whose boundary is the 10 Dimensional manifold
$\partial D_{11}=M_{10}$. It is independent of the particular $D_{11}$
because the integrand is closed due to the Bianchi identities which
are\begin{eqnarray}
dH_{3}=0,\, dF_{0} & = & 0,\, dF_{2}=F_{0}H_{3},\label{BiIIA1}\\
dF_{4} & = & F_{2}\wedge H_{3}.\label{BiIIA2}\end{eqnarray}
 The Bianchi identities are solved (locally) by\begin{eqnarray}
F_{2}=dC_{1}+F_{0}B_{2}, & F_{4}=dC_{3}-H_{3}\wedge C_{1}+\frac{F_{0}}{2}B_{2}\wedge B_{2},\label{BiIIAsoln1}\\
 & F_{0}=m_{0}={\rm const.}\label{BIIIAsoln2}\end{eqnarray}
 The gauge invariances of this system are as follows:\begin{equation}
\delta B_{2}=d\Lambda_{1},\,\delta C_{1}=d\Lambda_{0}-m_{0}\Lambda_{1},\,\delta C_{3}=d\Lambda_{2}-H_{3}\Lambda_{0}-m_{0}B_{2}\wedge\Lambda_{1}.\label{IIAguage}\end{equation}
 In the IIB case the branes that can be electrically coupled were
D3 and D1 branes. In IIA we can only couple D2 and D0 branes. In order
to couple higher branes we need to use the same trick as in the IIB
case. For the moment we will ignore eight branes and D-particles,
set $m_{0}=0$ and write down the effective terms for D6, D4 and D2
branes. These are,\begin{eqnarray}
I_{6} & = & \frac{\mu_{6}}{(2\pi)^{3}}\int_{W_{7}}[\frac{{\mathcal{F}}_{2}^{2}}{2!}\wedge C_{3}-\frac{{\mathcal{F}}_{2}^{3}}{3!}\wedge C_{1}],\label{I6}\\
I_{4} & = & \frac{\mu_{4}}{(2\pi)^{3}}\int_{W_{5}}[-{\mathcal{F}}_{2}\wedge C_{3}+\frac{{\mathcal{F}}_{2}^{2}}{2!}\wedge C_{1}],\label{I4}\\
I_{2} & = & \frac{\mu_{2}}{(2\pi)^{3}}\int_{W_{3}}\int[C_{3}-{\mathcal{F}}_{2}\wedge C_{1}],\label{I2}\end{eqnarray}
 with $\mu_{6}=\sqrt{2\pi}q,\,\mu_{4}=(2\pi)^{3/2}q,\,\mu_{2}=(2\pi)^{5/2}q,\, q=\pm1,0$.
As in the IIB case there is an anomaly in these actions that has to
be cancelled by inflow from the bulk\begin{equation}
\delta I_{6}=-\frac{\mu_{6}}{(2\pi)^{3}}\int_{W_{7}}\Lambda_{2}\wedge{\mathcal{F}}_{2}\wedge H_{3},\,\delta I_{4}=-\frac{\mu_{4}}{(2\pi)^{3}}\int_{W_{5}}\Lambda_{2}\wedge H_{3}.\label{deltaI64}\end{equation}
 Again there is at first sight an ambiguity in how the WZNW term can
be split up, but as in the IIB case consistency of the Bianchi identities
fixes this. The crucial identity is the one for $F_{2}$ which can
be fixed by T duality from the corresponding identity (\ref{F3BI})
whose validity was established in turn by the consistency of the IIB
equations with the self-duality constraint and anomaly inflow. Thus
we argue that \begin{equation}
dF_{2}=\mu_{6}\delta_{3}(M_{10}\rightarrow W_{7}).\label{F2Bi}\end{equation}

Note that as in the IIB case this is precisely what we would have
got if we just had a bulk action $\int F_{8}\wedge*F_{8}$ coupled
to $\int_{W_{7}}C_{7}$ but as in that case it is not possible to
have this in the IIA action given that we need also the lower rank
forms.

At least for trivial fluxes the topological term in (\ref{SIIA})
can be rewritten as a ten dimensional integral \[
-\frac{1}{2}\int_{M_{10}}C_{3}\wedge F_{4}\wedge H_{3}\]
 Then as in the IIB case the ambiguity gets fixed with $\lambda=2$
(see discussion between (\ref{Bimod}) and (\ref{mod2Bi2}) ) so that
the topological term gets split up as \begin{equation}
-\frac{1}{2}\int C_{3}\wedge F_{4}\wedge H_{3}\rightarrow-\int C_{3}\wedge F_{4}\wedge H_{3}+\frac{1}{2}\int C_{3}\wedge F'_{4}\wedge H_{3},\label{IIAsplit}\end{equation}
 where $dF'_{4}=F_{2}\wedge H_{3}$ and \begin{equation}
dF_{4}=\mu_{6}{\mathcal{F}}_{2}\delta_{3}(M_{10}\rightarrow W_{7})+\mu_{4}\delta_{5}(M_{10}\rightarrow W_{5})+F_{2}\wedge H_{3}\label{BiF4}\end{equation}
 It is consistency with $d^{2}=0$ and (\ref{F2Bi}) that enabled
us to fix $\lambda=2$ as in the IIB case. As before the source terms
in (\ref{BiF4}) give an anomaly in the gauge transformation of (\ref{IIAsplit})
which cancels the anomaly of the D6 and D4 brane actions (\ref{deltaI64}).

Let us now turn on $F_{0}$ flux. First let us discuss the corresponding
processes by constructing a consistent set of Bianchi identities.
Then we will investigate whether there is a bulk plus brane action
to describe them.

From the foregoing discussion the Bianchi identities in the presence
of six branes (for simplicity we'll not introduce 4- or 2-branes)
are changed from (\ref{BiIIA1})(\ref{BiIIA2}) to\begin{eqnarray}
dH_{3}=0, &  & dF_{2}=H_{3}F_{0}+\sum_{i}\mu_{6}^{i}\delta_{3}(M_{10}\rightarrow W_{7}^{i}),\label{BiIIA1mod}\\
dF_{4} & = & H_{3}\wedge F_{2}+\sum_{i}\mu_{6}^{i}{\mathcal{F}}_{2}\delta_{3}(M_{10}\rightarrow W_{7}^{i}).\label{BiIIA2mod}\end{eqnarray}

Note that we have now generalized to a set of six-branes. Suppose
the six-branes wrap the 3+1 external dimensions and a three cycle
($\alpha$) in the Calabi-Yau manifold $X$. Integrating the second
of (\ref{BiIIA1mod}) over the dual cycle $\beta$ we get (since $F_{2}$
is globally defined)\begin{equation}
0=\int_{\beta}dF_{2}=m_{0}\int_{\beta}H_{3}+\sum_{i}\mu_{6}\label{betaInt}\end{equation}
 This relation was observed in \cite{DeWolfe:2005uu}. It tells us
that if a net number of six branes wrapping a three cycle are present
then we need to have non-zero $F_{0}$ flux and also put $H_{3}$
flux through the dual cycle. Given the flux quantization\begin{equation}
\int_{\beta}H_{3}=2\pi p,\, p\epsilon{\mathcal{Z}}\label{HfluxIIA}\end{equation}
 we get \begin{equation}
2\pi m_{0}p=\sqrt{2\pi}N\label{m0p}\end{equation}
 Note that this relation is only consistent if $m_{0}$ is quantized
in units of $1/\sqrt{2\pi}$ a fact that can be established independently
by coupling eight branes as we shall see later. In the context of
type I' theory a similar result was observed in \cite{Polchinski:1995df}.

The interesting processes are those which changes the ten-form flux
$m_{0}$. These would change the ten dimensional cosmological constant
and are caused by nucleated eight-branes. A consistent set of Bianchi
identities in the presence of both eight and six branes is (henceforth
the wedge product symbol should be understood from the context)\begin{eqnarray}
dF_{0} & = & \mu_{8}\delta_{1},\label{BiF0}\\
dF_{2} & = & F_{0}H_{3}+\mu_{8}{\mathcal{F}}_{2}\delta_{1}+\mu_{6}\delta_{3},\label{BiF2}\\
dF_{4} & = & F_{2}H_{3}+\mu_{8}\frac{{\mathcal{F}}_{2}^{2}}{2!}\delta_{1}+\mu_{6}{\mathcal{F}}_{2}\delta_{3}+\mu_{4}\delta_{5}.\label{BIF4}\end{eqnarray}
 In the above $\delta_{1}=\delta_{1}(M_{10}\rightarrow W_{9}),\,\delta_{3}=\delta_{3}(M_{10}\rightarrow W_{7}),\,\delta_{5}=\delta_{5}(M_{10}\rightarrow W_{5}),$
where the $W_{i}$ are the world volumes of the 8-, 6-, and 4-branes.
Also $\mu_{8}=q/\sqrt{2\pi},\, q=\pm1,0$.

The consistency of these Bianchi identities can be easily checked
by operating with the exterior derivative and using $d{\mathcal{F}}_{2}=H_{3}$.
It should be noted that the existence of the eight-brane, signaled
by the non-zero RHS of (\ref{BiF0}) necessitates the eight-brane
terms in the other two Bianchi identities. In passing we note that
the system cannot accommodate a NS five brane since an equation of
the form\[
dH_{3}=\mu_{5}\delta_{4}(M_{10}\rightarrow W_{6})\]
 would be incompatible with (\ref{BiF2})(\ref{BIF4}) as the above
consistency check depended on having $dH_{3}=0$.

Let us now consider the concrete framework for moduli stabilization
in \cite{DeWolfe:2005uu}. This has D6 branes O6 planes and $F_{0},\,*F_{4}$
and $H_{3}$ flux. First consider the nucleation of an eight brane
wrapping a two sphere $S_{2}$ in non-compact space and the whole
Calabi-Yau $X$ in this set up. Integrating (\ref{BiF0}) over a line
$R$ going from inside to the outside of the $S_{2}$ in the radial
direction,\begin{equation}
\Delta F_{0}=m_{0}^{{\textrm{out}}}-m_{0}^{{\textrm{in}}}=\mu_{8}=\frac{1}{\sqrt{2\pi}}.\label{DeltaF0}\end{equation}
 This implies that $m_{0}$ is quantized in units of $1/\sqrt{2\pi}$
in agreement with our earlier argument. Integrating (\ref{BiF2})
over the the three cycle $\beta$ that is dual to that wrapped by
the six brane we get (after putting $m_{0}=m/\sqrt{2\pi},\, m\varepsilon{\mathcal{Z}}$)
equation (\ref{m0p}) rewritten as \begin{equation}
mp+N=0\label{mpN}\end{equation}
 where as before $N=N_{D6}-N_{O6}$. In crossing the nucleated 8-brane
domain wall $m\rightarrow m\pm1$, so in order to maintain the above
equation the number of D6 branes must change,\begin{equation}
N\rightarrow N\mp p.\label{D6change}\end{equation}
 In attempts to create standard-like models from type IIA the gauge
group comes from a stack of D6-branes. The above process will result
in changing the gauge group.

While this discussion shows that once six or eight branes are nucleated
various interesting processes (which change fluxes and the number
of branes) can take place, the nucleation itself is a quantum process
and at the very least one should expect that there be an action that
is able to describe this process. In the IIB case we argued that there
is no consistent action essentially because of the self-duality problem
even if we use the anomaly inflow argument. Above we showed that in
IIA in the absence of $F_{0}$ flux it is possible to get an action
for six branes magnetically coupled to the bulk action by using anomaly
inflow from the bulk to the brane. However as we've discussed above
flux stabilization of moduli in IIA requires us to turn on $F_{0}$
flux. Furthermore nucleation of eight branes will cause this flux
to jump. So it is imperative to have an action that incorporates eight
branes as well as six and four branes in the presence of non-zero
$F_{0}$ flux. Below we will try to construct such an action.

The main problem is to write the topological term in the bulk action
in such a way that, in the presence of sources, its anomalies will
cancel the anomaly of the magnetically coupled higher dimensional
branes, where the latter are written only in terms of the fields in
the bulk action. The term in question is (see (\ref{SIIA})) \[
S_{top}=-\frac{1}{2(2\pi)^{3}}\int_{D_{11}}H_{3}F_{4}F_{4}.\]

As we mentioned earlier this term is well defined in the absence of
magnetically coupled branes. In the presence of such sources however
the integrand is not closed so we have to find a way to define this
term properly. In order to do so we will first reexpress this term
(in the absence of magnetic couplings) as a ten dimensional integral
by writing

\begin{equation}
F_{4}=dC_{3}-H_{3}C_{1}+F_{0}^{b}\frac{B_{2}^{2}}{2!}+F_{4}^{b},\, dF_{4}^{b}=0,\, F_{2}=dC_{1},\, H_{3}=dB_{2}+H_{3}^{b}\label{BIsolns}\end{equation}
 Here we have included a background flux term for the four form (and
we've renamed the background zero form flux $F_{0}^{b}\equiv m_{0}^{out}$)
and NS three form flux. The stabilization discussed in \cite{DeWolfe:2005uu}
actually requires both four form and NS three form flux but if we
have both there will be extra (closed) terms which are necessarily
eleven dimensional but will not affect the argument below. With the
above we can write (after some algebra) the topological term as\begin{eqnarray}
S_{top} & = & -\frac{1}{2(2\pi)^{3}}\int_{D_{11}}H_{3}F_{4}F_{4}=-\frac{1}{(2\pi)^{3}}\int_{M_{10}}[\frac{1}{2}C_{3}H_{3}(F_{4}+F_{4}^{b})+\frac{1}{2}B_{2}F_{4}^{b}F_{4}^{b}\label{Stop}\\
 & - & \frac{1}{2}C_{3}H_{3}F_{0}^{b}\frac{B_{2}^{2}}{2!}-F_{4}F_{0}^{b}\frac{B_{2}^{3}}{3!}+F_{2}F_{0}^{b}\frac{B_{2}^{4}}{4!}-2(F_{0}^{b})^{2}\frac{B_{2}^{5}}{5!}+\ldots]\nonumber \end{eqnarray}
 The ellipses indicate terms having a factor of $H_{3}^{b}$ which
will not be changed in the presence of magnetic couplings to D-branes,
so that we do not need their explicit expression for what follows.
The gauge invariant expression for the coupling of RR fields to a
p- brane is \cite{Green:1996bh}\begin{equation}
I=\frac{\mu_{p}}{(2\pi)^{3}}\int_{W_{p+1}}\{{\textbf{C}}e^{-(f_{2}+B_{2})}+F_{0}\sum_{r}\frac{(-1)^{r}}{(r+1)!}\omega_{2r+1}(f_{2},a)\}.\label{Itotal}\end{equation}
 Here ${\textbf{C}}$ is the formal sum of all the RR fields and $\omega$
is the Chern-Simons form for the gauge field $a$. We are of course
ignoring the gravitational anomaly contribution which is irrelevant
to our discussion. In addition to the usual RR and NS transformations
this action is invariant under the additional NSNS transformations
that are present (see Appendix A) when there is constant zero form
flux ($dF_{0}=0,\, F_{0}\ne0$) \begin{equation}
\delta{\bf C}=-F_{0}e^{B_{2}}\Lambda_{1}.\label{Ctotvar}\end{equation}
 As before since the bulk action does not contain the higher form
fields we need to use the anomaly inflow mechanism to get a gauge
invariant bulk plus brane action. Let us see how this works for the
coupling of the D8 brane. Since the other branes are contained in
this one it is sufficient to demonstrate the mechanism for this case.
The first step is to rewrite (as with the other cases) the D8 action
by removing the terms which are dependent on gauge fields of rank
higher than four since these are not contained in the bulk action.
The terms which are removed are,\begin{equation}
\Delta I_{8}=\frac{\mu_{8}}{(2\pi)^{3}}\int_{W_{9}}[C_{9}-C_{7}{\cal F}_{2}+C_{5}\frac{{\cal F}_{2}}{2!}]\label{DeltaI8}\end{equation}
 Under RR and NSNS gauge transformations we have using (see Appendix
A) \begin{equation}
\delta C_{9}=d\Lambda_{8}-H_{3}\Lambda_{6}-F_{0}^{b}\frac{B^{4}}{4!}\Lambda_{1},\,{\rm etc.}\label{deltaC}\end{equation}
 (and after some cancellations) the anomaly that needs to flow in
from the bulk,\begin{equation}
\delta\Delta I_{8}=\frac{\mu_{8}}{(2\pi)^{3}}\int[-H_{3}\Lambda_{2}\frac{{\cal F}_{2}^{2}}{2!}-F_{0}^{b}(\frac{B^{4}}{4!}-\frac{B^{3}}{3!}{\cal F}_{2}+\frac{B^{2}}{2!}\frac{{\cal F}_{2}^{2}}{2!})\Lambda_{1}].\label{deltaI8}\end{equation}
 Now the question is whether the bulk topological term can provide
this inflow. As we discussed before the topological term (written
as an 11-dimensional integral over a disc) is well defined only if
the integrand is closed. This is so because $H_{3}$ is closed and
the RR field strengths are $d_{H}$ closed. In the presence of magnetic
sources (i.e. 4, 6 or 8-branes) however the RR field strengths are
not $d_{H}$ closed. The question is whether the resulting ambiguity
can be resolved so that the anomaly of the truncated D-brane action
is cancelled.

The unique way of doing this is to rewrite the topological term (in
its ten dimensional form (\ref{Stop}) as follows (ignoring terms
involving $H_{3}^{b}$):\begin{eqnarray}
S_{top} & = & \frac{1}{(2\pi)^{3}}\int_{M_{10}}[-C_{3}H_{3}F_{4}+\frac{1}{2}C_{3}H_{3}F'_{4}-\frac{1}{2}B_{2}F_{4}^{b}F_{4}^{b}\label{Stop2}\\
 &  & +\frac{1}{2}C_{3}H_{3}F_{0}^{b}\frac{B_{2}^{2}}{2!}-F_{4}F_{0}^{b}\frac{B_{2}^{3}}{3!}+F_{2}F_{0}^{b}\frac{B_{2}^{4}}{4!}+(F_{0}^{b})^{2}\frac{B_{2}^{5}}{5!}+F_{0}^{b}F_{0}\frac{B_{2}^{5}}{5!}]\nonumber \end{eqnarray}
 $F_{4}^{b},F_{0}^{b}$, are taken to be the background values of
these fluxes at infinity (i.e. outside the nucleated D8-brane) and
$F_{4}'$ is defined as $F'_{4}=dC_{3}-H_{3}C_{1}+F_{0}^{b}\frac{B_{2}^{2}}{2!}+F_{4}^{b}$
(see (\ref{BIsolns})). $F_{4},$$F_{2}$ and $F_{0}$ satisfy the
Bianchi identities \begin{eqnarray}
dF_{0}=\mu_{8}\delta_{1}, &  & dF_{2}-F_{0}H_{3}=\mu_{8}{\cal F}_{2}\delta_{1}\label{BI802}\\
dF_{4}-F_{2}H_{3} & = & \mu_{8}\frac{{\cal F}_{2}^{2}}{2!}\delta_{1}\label{BI84}\end{eqnarray}
 So as was the case in IIB (see discussion after (\ref{stopIIB})
we may write (defining $\theta_{i},\, i=0,2,5$ ) to be coexact solutions
of the three Bianchi identities (\ref{BI802},\ref{BI84}) \begin{equation}
F_{0}=F_{0}^{b}+\theta_{0},\, F_{2}=dC_{1}+\theta_{2},\, F_{4}=F_{4}'+\theta_{4}.\label{F4F2F0}\end{equation}
 $ $The gauge transformations are given in (\ref{IIAguage}). The
non-zero right hand sides of these equations imply that there is an
anomaly in the bulk topological term (\ref{Stop2}) given by\begin{eqnarray}
\delta_{anom}S_{top} & = & \frac{1}{(2\pi)^{3}}\int_{M_{10}}[-\Lambda_{2}H_{3}dF_{4}+dF_{4}F_{0}^{b}\frac{B_{2}^{2}}{2!}\Lambda_{1}-dF_{2}F_{0}^{b}\frac{B_{2}^{3}}{3!}\Lambda_{1}-dF_{0}F_{0}^{b}\frac{B_{2}^{4}}{4!}\Lambda_{1}]|_{anom}\nonumber \\
 & = & \frac{\mu_{8}}{(2\pi)^{3}}\int_{M_{10}}[-H_{3}\Lambda_{2}\frac{{\cal F}_{2}^{2}}{2!}-F_{0}^{b}\frac{{\cal F}_{2}^{2}}{2!}\frac{B_{2}^{2}}{2!}\Lambda_{1}+F_{0}^{b}{\cal F}_{2}\frac{B_{2}^{3}}{3!}\Lambda_{1}-F_{0}^{b}\frac{B_{2}^{4}}{4!}\Lambda_{1}]\delta_{1}.\label{deltaanomStop}\end{eqnarray}
 The delta function at the end just restricts the integral to the
eight-brane world volume so that this is exactly equal to (\ref{deltaI8}).
In other words the anomaly in the source action coming from restricting
the gauge fields to those that are present in the bulk lagrangian,
is cancelled by anomaly inflow from the bulk.

In spite of the cancellation of anomalies that we have demonstrated
above, the action (in particular the topological term (\ref{Stop2}))
is not written in a background independent way. It is not clear that
it is possible to do this in a Lorentz invariant fashion. This of
course was the problem in the IIB case as well. However in the IIA
case we do have an action with the bulk coupled to the higher dimensional
branes that can be used to compute the quantum fluctuations around
a given background %
\footnote{In Appendix B we discuss an elegant geometrical formulation due to
Belov and Moore \cite{Belov:2006} where considerable progress towards
achieving a background independent formulation is made.%
}.

\section{Conclusions}

In this paper we have looked at the dynamics of transitions between
different flux vacua (for a given Calabi-Yau compactification). Our
arguments imply that even though there are classical configurations
that can divide the universe into different regions, with different
numbers of branes and flux values, it is difficult to find a Lorentz
invariant background independent supergravity action that includes
the bulk theory and the nucleated higher dimensional branes. This
means that it is hard to see how to describe the quantum mechanical
process that causes transitions between such regions. In other words
it is not entirely clear that there is no superselection rule that
forbids such transitions so that each point of the landscape is simply
a different sector isolated from the rest. If this is true there is
no need to assign reality to the whole landscape. In other words one
might think that finding a standard model with a nearly zero cosmological
constant etc. is really nothing more than just fitting the data. Philosophically
this would be no different from what particle physicists were doing
in the 70's and 80's i.e. GUT model building.

On the other hand unlike in these field theories, here there are classical
configurations where the universe is separated by domain walls into
regions with different values of the fundamental constants number
of generations etc. So it is conceivable that even though within the
low-energy approximation we have not been able to find a quantum description
of this brane nucleation process, in that a background independent
Lorentz invariant supergravity action does not seem to exist, the
full string theory may still admit such processes. For instance it
is likely that, since the brane actions that we have been using are
really valid effective descriptions only at scales which are long
compared to the string scale, and the gauge invariance problems that
we have highlighted occur at the locations of the branes, a proper
string theoretic description of the brane (for instance as a soliton
in a string field theory) should automatically take care of this problem.

\section{Acknowledgments}

I wish to thank Rob Myers for very useful discussions and Oliver DeWolfe
for stimulating comments on the manuscript. I also wish to thank Dima
Belov and Greg Moore for discussions (and the former also for correspondence
about their work \cite{Belov:2006}) that stimulated this revised
version. In addition I wish to thank Ben Shlaer for discussions. Finally
I wish to thank the Aspen center for physics and KITP Santa Barbara
for hospitality while this paper was being revised. This work is supported
by DOE grant No. DE-FG02-91-ER-40672 and the Perimeter Institute.

\section*{Appendix A: The dual form of the IIA action}

In this appendix we will first discuss a formal self-dual action and
then a dual form of the IIA action that might be suitable for coupling
to the higher dimensional branes \cite{Bergshoeff:2001pv}. Define
(for type IIA) the formal sum of forms

\begin{equation}
{\textbf{A}}=\sum_{n=1}^{5}A_{2n-1},\,{\textbf{F}}=\sum_{n=0}^{5}F_{2n}\label{formsum}\end{equation}

with a similar sum over even (odd) rank gauge fields (field strengths)
for type IIB. ${\bf F}$ satisfies the Bianchi identity

\[
d{\bf F}=H_{3}{\bf F}\]
 The usual form of the gauge fields (used in the text) is obtained
by substituting ${\textbf{A}}=e^{-B_{2}}{\textbf{C}}$. Also (locally)
we solve the Bianchi identities by \begin{equation}
{\bf F}=(d{\bf A}+F_{0})e^{B}=d{\bf C}-H_{3}{\bf C}+F_{0}e^{B_{2}}.\label{Ffieldstrength}\end{equation}

The action is \begin{equation}
(2\pi)^{3}S=-\frac{1}{2}\int_{M_{10}}(|{\bf F}|^{2}+|H_{3}|^{2})\label{selfdualactn}\end{equation}

The RR gauge transformations are $\delta{\bf A}=d\tilde{{\bf \Lambda}}$,
($\tilde{{\bf \Lambda}}=\sum_{n=0}^{4}\tilde{\Lambda}_{2n}$) and
the NSNS gauge transformations are\begin{eqnarray*}
\delta B_{2} & = & d\Lambda_{1},\,\delta{\bf A}=-F_{0}\Lambda_{1}-d\Lambda_{1}{\bf A}\end{eqnarray*}
 The corresponding gauge transformations for ${\bf C}$ are (with
${\bf }$${\bf \Lambda}=e^{B_{2}}\tilde{{\bf \Lambda}}$)\begin{eqnarray}
\delta_{RR}{\bf C} & = & d{\bf \Lambda}-H_{3}{\bf \Lambda}\label{CRR}\\
\delta_{NS}{\bf C} & = & -F_{0}e^{B_{2}}\Lambda_{1}\label{CNS}\end{eqnarray}
 \[
\]
 One may couple the branes to this as in (\ref{Itotal}). This is
of course gauge invariant (provided $dF_{0}=0$). However this bulk
plus brane action has too many degrees of freedom since it has both
electric and magnetic terms which are dual to each other. To cut them
down we need to impose the self duality condition\[
{\bf F}=(-1)^{n}*{\bf F}\]
 at the level of the equations of motion. Note that we cannot put
this in the action even for type IIA since if we did so there would
be no dynamics for the gauge field ${\bf A}$ (see (\ref{Ffieldstrength})
- in IIB of course this term would vanish because the wedge product
of an odd form with itself vanishes). As observed by the authors of
\cite{Bergshoeff:2001pv} this is not a proper action for the quantum
theory since the latter must propagate only the physical degrees of
freedom.

A suggestion for such an action in \cite{Bergshoeff:2001pv} is, \begin{eqnarray}
(2\pi)^{3}S_{IIA} & = & -\frac{1}{2}\int_{M_{10}}\{(|H_{3}|^{2}+|F_{0}|^{2}+|F_{2}|^{2}+|F_{4}|^{2})-F_{4}F_{4}B_{2}\nonumber \\
 & + & F_{4}F_{2}B_{2}^{2}-\frac{1}{3}F_{2}^{2}B_{2}^{2}-\frac{1}{3}F_{0}F_{4}B_{2}^{3}+\frac{1}{4}F_{0}F_{2}B_{2}^{4}-\frac{1}{20}F_{0}^{2}B_{2}^{5}\nonumber \\
 & - & 2F_{0}dA_{9}+2(F_{2}-B_{2}F_{0})dA_{7}-2(F_{4}-B_{2}F_{2}+\frac{1}{2}B_{2}^{2}F_{0})dA_{5}\}\label{BIIA}\end{eqnarray}
 Here the field strength $H_{3}=dB_{2},$ but the RR field strengths
are taken to be independent {}``black box'' fields. The Lagrange
multiplier fields $A_{i}$ are designed to enforce the Bianchi identities
for these RR fields but they are also the fields which are supposed
to be sourced by the higher dimensional branes. Under RR gauge transformations
$F_{i}$ are invariant and \[
\delta A_{9}=d\Lambda_{8},\,\delta A_{7}=d\Lambda_{6},\,\delta A_{5}=d\Lambda_{4}.\]
 The NSNS gauge transformations (and supersymmetry transformations)
are realized only on the formal sum of all the form fields.

This is however not a suitable action for our purposes. For instance
If we couple an eight-brane the WZNW part of the action will be\begin{equation}
I_{8}=\mu_{8}\int_{W_{9}}(e^{-f_{2}}{\textbf{A}}-\frac{f^{4}}{4!}aF_{0})=\mu_{8}\int_{W_{9}}(A_{9}-A_{7}f_{2}+A_{5}\frac{f_{2}^{2}}{2!}-A_{3}\frac{f_{2}^{3}}{3!}+\ldots)\label{I8Abasis}\end{equation}
 where as before $f_{2}=da$ is the world volume gauge field strength
which transforms as $\delta f_{2}=-d\Lambda_{1}$, under NSNS gauge
transformations. Note that this field cannot be set to zero in string
theory. The important point here is that gauge invariance necessitates
the presence of all the lower dimensional branes. The bulk action
however is independent of the corresponding fields $A_{3}$ and $A_{1}$
and so variation with respect to these will only be consistent with
the absence of eight branes. The same is the case for six-branes and
four-branes. Thus this form of the action is not really suitable for
coupling higher dimensional branes.

\section*{Appendix B: A background independent action for IIA}

In this appendix we discuss the work of Belov and Moore \cite{Belov:2006}
- though in a mathematically unsophisticated fashion. Define (for
type IIA) the formal sum of forms as in the previous appendix:

\begin{equation}
{\textbf{C}}=\sum_{i=1}^{5}C_{2n-1},\,{\textbf{F}}=\sum_{i=0}^{5}F_{2n}\label{formsum2}\end{equation}

Write the bulk action for the RR gauge fields as (again setting the
dilaton to zero) \cite{Moore:2002cp,Diaconescu:2000wy}, \cite{Belov:2006}
\begin{eqnarray}
(2\pi)^{3}S & = & -\frac{1}{2}\int_{M_{10}}(F_{0}*F_{0}+F_{2}*F_{2}+F_{4}*F_{4})\label{Kterms}\\
 &  & -\frac{1}{2}\int_{M_{10}}(F_{0}F_{10}-F_{2}F_{8}+F_{4}F_{6}).\label{WZterms}\end{eqnarray}

Here all products are wedge products and the field strengths are required
to obey the Bianchi identities\begin{equation}
d{\bf F}-H_{3}{\bf F}\equiv d_{H}{\bf F}=0.\label{BI}\end{equation}
 Locally, in the absence of sources, we have the solution (\ref{Ffieldstrength}).

As it stands the above action is not well defined without choosing
a Lagrangian subspace of the space of gauge fields which tells us
how $F_{6},F_{8},F_{10}$ are related to $F_{0},F_{2},F_{4}$. However
in the absence of sources it is completely equivalent to the standard
form of the action (\ref{SIIA}). This is because the topological
term in the action (with $M_{10}=\partial D_{11}$) can be written
as\begin{eqnarray}
(2\pi)^{3}S_{top}=-\frac{1}{2}\int_{D_{11}}d(F_{0}F_{10}-F_{2}F_{8}+F_{4}F_{6})\nonumber \\
=-\frac{1}{2}\int_{D_{11}}(F_{0}H_{3}F_{8}-H_{3}F_{0}F_{8}-F_{2}H_{3}F_{6} & +H_{3}F_{2}F_{6}+F_{4}H_{3}F_{4})\nonumber \\
=-\frac{1}{2}\int_{D_{11}}H_{3}F_{4}F_{4}.\label{wz2}\end{eqnarray}

Here to get the second line we used the Bianchi identities. This is
the usual form of the topological term. In the presence of sources
the last expression is not well defined by itself since the integrand
is not closed. The starting point (\ref{WZterms}) is well-defined
provided a choice of a Lagrangian submanifold is made. Now Belov and
Moore introduce a {}``trivialization'' of the D-brane current. Let
us pursue an alternate strategy which illustrates the problems of
trying to retain the usual formulation of the action. Doing the above
calculation, using now the Bianchi identities with sources\begin{equation}
d_{H}{\bf F}={\bf j},\label{BIS}\end{equation}
 where \[
{\bf j}=\sum_{r=1}^{5}j_{2r-1},\]
 we get the following additional terms in the last two lines of (\ref{wz2})
:\begin{equation}
-\frac{1}{2}\int_{D_{11}}(F_{10}j_{1}-F_{8}j_{3}+F_{6}j_{5}+F_{4}j_{7}-F_{2}j_{9}).\label{WZS}\end{equation}
 Note that since $d_{H}^{2}=0$ the sources must satisfy $d_{H}{\bf j}=0.$
Explicitly we have the following expressions for these sources\begin{eqnarray}
j_{1} & = & \mu_{8}\delta_{1},\, j_{3}=\mu_{8}{\cal F}_{2}\delta_{1}-\mu_{6}\delta_{3},\, j_{5}=\mu_{8}\frac{{\cal F}_{2}^{2}}{2!}\delta_{1}-\mu_{6}{\cal F}_{2}\delta_{3}+\mu_{4}\delta_{5}\nonumber \\
j_{7} & = & \mu_{8}\frac{{\cal F}_{2}^{3}}{3!}\delta_{1}-\mu_{6}\frac{{\cal F}_{2}^{2}}{2!}\delta_{3}+\mu_{4}{\cal F}_{2}\delta_{5}-\mu_{2}\delta_{7}\nonumber \\
j_{9} & = & \mu_{8}\frac{{\cal F}_{2}^{4}}{4!}\delta_{1}-\mu_{6}\frac{{\cal F}_{2}^{3}}{3!}\delta_{3}+\mu_{4}\frac{{\cal F}_{2}^{2}}{2!}\delta_{5}-\mu_{2}{\cal F}_{2}\delta_{7}+\mu_{0}\delta_{9}\label{sources}\end{eqnarray}
 where $\mu_{p}=(2\pi)^{\frac{7-p}{2}}$, and ${\cal F}_{2}=B_{2}+f_{2}$,
with $f_{2}=da_{1}$ where $a_{1}$ is the world volume gauge field
of D-branes which transforms as $\delta f_{2}=-d\Lambda_{1}$, under
NSNS gauge transformations. Also $\delta_{2r-1}$ is a delta function
$2r-1$ form which has support on the $11-2r$ dimensional world volume
of a $10-2r$ dimensional D-brane. The source action may then be written
as\begin{equation}
(2\pi)^{3}I=\sum_{r=0}^{4}\mu_{2r}\int_{W_{2r+1}}({\bf C}e^{-{\cal F}_{2}}-F_{0}\frac{(-f)^{r}}{r+1!}a)=\int_{M_{10}}({\bf C}\bar{j}-F_{0}\sum_{r=0}^{4}\frac{(-f)^{r}}{(r+1)!}a{\bf \delta}|_{9-2r})\label{Localactn}\end{equation}
 where ${\bf \delta}=\sum_{r=0}^{4}\mu_{8-2r}\delta_{2r+1}$ and $\bar{j}=\sum_{r=0}(-1)^{r}j_{2r+1}$.
From the fact that locally on the brane we have (\ref{Ffieldstrength})
and that $\bar{j}$ satisfies the identity\begin{equation}
\bar{d}\bar{j}\equiv d\bar{j}+H_{3}\bar{j}=0\label{dbarjbar}\end{equation}
 we get $d(C\bar{j})=F\bar{j}-C\bar{d}\bar{j}-F_{0}e^{B}\bar{j}=F\bar{j}-F_{0}e^{B}\bar{j}$.
Using also the explicit form of $\bar{j}$ (with the defining formal
sum extended to 11 dimensions ($r=5)$ with $j_{11}=(e^{-{\cal F}_{2}}\delta)_{11}$
) we get\begin{equation}
I=\int_{D_{11}}{\bf F}\bar{j}\label{ID}\end{equation}

Now if we add these source terms with half strength to the bulk action
we have from (\ref{WZS})\begin{equation}
(2\pi)^{3}(S_{WZ}+\frac{1}{2}I)=-\frac{1}{2}\int_{D_{11}}H_{3}F_{4}F_{4}-\int_{D_{11}}(F_{4}j_{7}-F_{2}j_{9}-F_{0}j_{11})\label{S+I}\end{equation}
 Thus the higher form fields have disappeared since the source terms
are truncated (as in the anomaly inflow argument). The sum is of course
manifestly gauge invariant and well defined (once a Lagrangian subspace
has been chosen \cite{Belov:2006}) since we started from a ten-dimensional
action which just depended on the curvatures, to which the local gauge
invariant action (\ref{Localactn}) was added. However some comments
are in order. Firstly the source terms cannot now be written as world
sheet integrals anymore unless the coupling of the higher branes (D4,
D6 and D8) is turned off. Thus if $\mu_{8}=\mu_{4}=\mu_{2}=0$ we
have for the second term in (\ref{S+I})\begin{eqnarray*}
\int_{D_{11}}[\mu_{2}(F_{4}-F_{2}{\cal F}_{2}+F_{0}\frac{{\cal F}_{2}^{2}}{2!})\delta_{7} & + & \mu_{0}(F_{2}-{\cal F}_{2}F_{0})\delta_{9}]\\
=\int_{M_{10}}[\mu_{2}(C_{3}-{\cal F}_{2}C_{1}+F_{0}\frac{f_{2}a_{1}}{2!})\delta_{7} & + & \mu_{0}(C_{1}-F_{0}a)\delta_{9}]\\
=\int_{W_{3}}(C_{3}-{\cal F}_{2}C_{1}+F_{0}\frac{f_{2}a}{2!}) & + & \int_{W_{1}}(C_{1}-F_{0}a)\end{eqnarray*}
 But there is no corresponding world volume integral representation
of the higher dimensional branes. This appears to be the price that
has to be paid to have an action which is independent of the base
point in contrast to our previous discussion which had Dirac string
singularities. In other words one can as in the text introduce particular
solutions of $d_{H}{\bf F}={\bf J}$ (trivializations of ${\bf J}$)
to rewrite the topological terms as integrals over the ten manifold
$M_{10}$. This is what is done in \cite{Belov:2006} - though in
a very elegant geometrical fashion. They have also argued that the
action is in some sense independent of this trivialization.

It should be remarked that although the topological terms for the
branes have been added with half strength (\ref{S+I}) the DBI term
must be added with full strength. It is only then that the usual (kappa
invariant BPS) form of the brane action is obtained when coupled to
the usual form of the bulk action with lower dimensional branes. In
other words the bulk action \eqref{Kterms}\eqref{WZterms} needs
to be coupled to a brane action which is not manifestly BPS in that
the charge is apparently half the tension. This is a reflection of
the fact that some of the brane topological terms are hidden in the
bulk topological term.

The above discussion shows that it is not possible to write the standard
form of the background independent bulk action with a coupling to
higher dimensional branes unless the latter are written in a non-local
form. As shown above for the lower dimensional electrically coupled
branes it is possible to rewrite them in the local form, but not for
the magnetically coupled branes. Of course the bulk action in the
form \eqref{Kterms}\eqref{WZterms} coupled to the local action \eqref{Localactn}
is indeed in a background independent form, but it is not properly
defined unless a Lagrangian submanifold is chosen \cite{Belov:2006}.
It is not clear that this can be done in a Lorentz invariant and background
independent fashion.

\bibliographystyle{apsrev} \bibliographystyle{apsrev}
\bibliography{myrefs}

\end{document}